\documentclass[12pt]{article}
 \usepackage{graphicx,amsfonts,amsmath}

 \renewcommand{\title}[1] {%
 \begingroup\begin{center}\vspace{0.0cm}\bf\Large
 \addtolength{\baselineskip}{1mm} #1 \end{center}\endgroup}

 \renewcommand{\author}[1] {%
 \begingroup\begin{center}\vspace{0.2cm}\bf #1 \vspace{0.2cm}
 \end{center}\endgroup}

  \newcommand{\address}[1] {%
 \begingroup\begin{center} #1 \end{center}\endgroup}

\newcommand\rv{{\mathbf {r}}}
\newcommand\pv{{\mathbf {p}}}
\newcommand\qv{{\mathbf {q}}}
\newcommand\M{{\mathfrak{M}}}
\newcommand\HC{{\mathcal{H}}}
\newcommand\F{{\mathcal{F}}}
\newcommand\Pd{\tfrac{1}{2}}
\newcommand\e{{\mathrm{e}}}
\newcommand\Pf{{\mathrm{Pf\,}}}
\newcommand\const{{\mathrm{const}}}

\begin{document}
 \title{Magnetic susceptibility of the 2D Ising model\\ on a finite
 lattice }{}%
 \author{A. I. Bugrij$^{\; *}$, O. Lisovyy$^{\;*,\;\dag}$}
 \address{
  $^{*\;}$Bogolyubov Institute for Theoretical Physics \\
  Metrolohichna str., 14-b, Kyiv-143, 03143, Ukraine \vspace{0.2cm} \\
  $^{\dag\;}$ Laboratoire de Math\'ematiques et Physique Th\'eorique CNRS/UMR 6083,\\
  Universit\'e de Tours, Parc de Grandmont, 37200 Tours, France}
  \date{}
 \begin{abstract}
 Form factor representation of the correlation function of the 2D Ising model
  on a cylinder is generalized to the case of
 arbitrary disposition of correlating spins. The magnetic
 susceptibility on a lattice, one of whose dimensions ($N$) is
 finite, is calculated in both para- and ferromagnetic regions
 of parameters of the model. The singularity structure of the susceptibility in the
 complex temperature plane at finite values of $N$ and the thermodynamic limit
 $N\rightarrow\infty$ are discussed.
 \end{abstract}

 \section{Introduction}
 The Ising model has long been a subject of great interest.
 The computation of the free energy \cite{Onsager} and
 spontaneous magnetization \cite{Yang}, Toeplitz determinant representations \cite{Montroll},
 form factor expansions \cite{Wu} and
 nonlinear differential equations \cite{tracy_mccoy,Jimbo} for the correlation
 functions are among the most important advances of the modern mathematical
 physics. The partition function of the 2D Ising model in zero
 field was calculated exactly
 \cite{McCoyWu} not only in the thermodynamic limit but also for finite
 lattices with different boundary conditions. The simplicity  of
 the corresponding expressions enables one to get an idea
 about the mechanism of appearance of critical singularities
 in thermodynamic quantities from both mathematical and physical
 points of view.

 Analytical expressions for thermodynamic quantities, which
 contain the dependence on lattice size, have numerous
 applications. For example, in computer simulation of
 thermodynamic systems or quantum field models one often needs such
 expressions to estimate the number of  degrees of freedom for
 which a discrete numerical model is adequate to the initial
 continuous and infinite system. It is worth mentioning that
 modern experiments and technologies often deal with finite-size
 systems. Theoretical analysis of such systems experiences
 the lack of exactly solvable examples.

 In this paper we present exact expressions for the 2-point
 correlation function and the susceptibility of the 2D Ising model
 on a lattice with one finite ($N=\const$) and the other infinite
 ($M\to\infty$)
 dimension. These expressions are very similar to well-known form factor
 expansions \cite{Palmev}, \cite{yamada1}.
 We investigate the singularity structure of the susceptibility for
 finite values of $N$ and discuss the thermodynamic limit $N\to\infty$.

\section{Correlation function $\langle\sigma(0,0)\sigma(x,0)\rangle$}
 The Ising model on  $M\times N$ square lattice (Fig. 1)
\begin{figure}[h] \begin{center}
 \includegraphics[height=80mm,keepaspectratio=true]
 {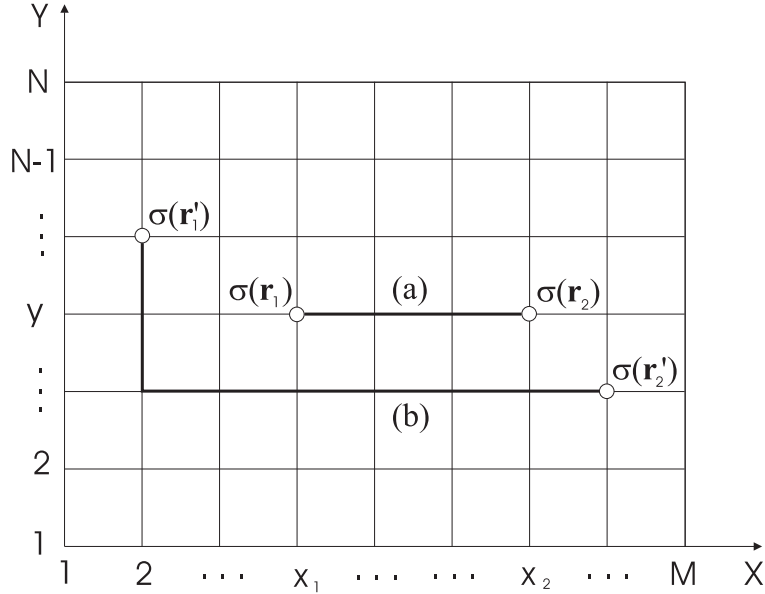}
 \caption{The numbering of the lattice sites and the variants of
 the
 disposition of correlating spins: a) along the cylinder axis,
 b) arbitrary disposition of spins on the lattice.}
 \end{center} \end{figure}
 is defined by the hamiltonian $H[\sigma]$ $$
H[\sigma]=-J\sum_{\rv}\sigma(\rv) (\nabla_x+\nabla_y)\sigma(\rv),
$$
 where the two-dimensional vector
$\rv=(x,y)$ labels the lattice sites: $x=1,2,\ldots,M$,
$y=1,2,\ldots,N$; the Ising spin $\sigma(\rv)$ in each site takes
on the values $\pm1$; the parameter $J>0$ defines the energy of
the coupling of adjacent spins. Shift operators $\nabla_x$,
$\nabla_y$ act as follows:
 $$
\nabla_x\sigma(x,y)=\sigma(x+1,y),\quad
\nabla_y\sigma(x,y)=\sigma(x,y+1). $$ Partition function  at the
temperature $\beta^{-1}$
\begin{equation}
\label{2.3} Z=\sum_{[\sigma]}\e^{-\beta H[\sigma]},
\end{equation}
and 2-point correlation function
\begin{equation}
 \label{2.4}
\langle\sigma(\rv_1)\sigma(\rv_2)\rangle=Z^{-1}\sum_{[\sigma]}\e^{-\beta
H[\sigma]}\sigma(\rv_1)\sigma(\rv_2)
\end{equation}
are given by the sums over all possible spin configurations. It is
convenient to introduce the following dimensionless parameters:
\begin{equation}
K=\beta J,\quad t=\tanh K, \quad s=\sinh 2K.
\end{equation}

 We will impose periodic boundary conditions on both axes.
 This gives two equations for the shift operators
 $\nabla_x$, $\nabla_y$: $$ (\nabla_x)^M=1,\quad (\nabla_y)^N=1. $$
 For such boundary conditions the partition function (\ref{2.3})
 can be written as a sum of four terms \cite{McCoyWu}:
\begin{equation}
\label{2.7} Z=(2\cosh^2 K)^{MN}\cdot
\frac{1}{2}\biggl(Q^{(f,f)}+Q^{(f,b)}+Q^{(b,f)}-Q^{(b,b)}\biggr).
\end{equation}
 Each of them is given by the pfaffian of the operator $\widehat{D}$
 (a lattice analogue of the Dirac operator)
\begin{equation}\label{2.8}
Q=\Pf\widehat{D},
\end{equation}
 defined by
 \begin{equation}\label{2.9}
 \widehat{D}=
 \left(\begin{array}{cccc}
 0 & 1 & 1+t\nabla_x & 1 \\
 -1 & 0 & 1 & 1+t\nabla_{y} \\
 -1-t\nabla_{-x} & -1 & 0 & 1 \\
 -1 & -1-t\nabla_{-y} & -1 & 0
 \end{array}\right).
 \end{equation}
 The upper indices
$(f,b)$ of the quantities $Q$  in (\ref{2.7}) correspond to
different types (antiperiodic or periodic) of boundary conditions
for the operators $\nabla_x,\nabla_y$ in (\ref{2.9}):
\begin{equation}
(\nabla_x^{(b)})^M=(\nabla_y^{(b)})^N=1,\quad
(\nabla_x^{(f)})^M=(\nabla_y^{(f)})^N=-1.
\end{equation}
When, for example, $M\gg N$ (i.e. the torus transforms into a
cylinder), then in the right hand side of (\ref{2.7}) only  the
``antiperiodic'' term survives:
\begin{equation}\label{2.11}
 Z=(2\cosh^2K)^{MN}Q^{(f,f)}.
 \end{equation}
Since the operator $\widehat{D}$ is translationally invariant, the
pfaffian (\ref{2.8}) can be easily computed. Using Fourier
transformation, one finds the following factorized representation
for the partition function (\ref{2.11}):
\begin{equation}\label{2.12}
Z=2^{MN}{\prod_{\qv}}^{(f,f)}(s^2+1-s\cdot\cos q_x-s\cdot\cos
q_y)^{1/2}.
\end{equation}
The superscript  $(f)$  in the  products (or sums below) implies
that the quasimomentum components $q_x$ and $q_y$ run in the
Brillouin zone over half-integer values in the units $2\pi/M$ and
$2\pi/N$, respectively; integer values correspond to the
superscript $(b)$. For example, $$
{\prod_{q_y}}^{(f)}\F(q_y)=\prod_{l=1}^N\F\biggl(\frac{2\pi}{N}(l+\Pd)
\biggr),\qquad  {\prod_{q_y}}^{(b)}
\F(q_y)=\prod_{l=1}^N\F\biggl(\frac{2\pi}{N}l\biggr). $$
 The product over one of the quasimomentum components in the right
 hand side of (\ref{2.12}) can be calculated in an explicit form,
 so that for the partition function one has
\begin{equation} \label{2.14}
Z=(2s)^{MN/2}{\prod_{q}}^{(f)}\e^{-M\gamma(q)/2}\bigl(1+
\e^{-M\gamma(q)}\bigr),
\end{equation}
 where the function $\gamma(q)$
 is the positive root of the equation
\begin{equation}
\sinh^2\frac{\gamma(q)}{2}=\sinh^2\frac{\mu}{2}+\sin^2\frac{q}{2},
\label{2.15}
\end{equation}
 and the parameter $\mu$ is the following function of $s$:
\begin{equation}
\sinh\frac{\mu}{2}=\frac{1}{\sqrt{2}}\bigl(\sqrt{s}-
1/\sqrt{s}\bigr).\label{2.16}
\end{equation}
 For $q\neq0$  $\gamma(q)$ remains positive in the whole range of
 the parameter $s$ ($0<s<\infty$), but $\gamma(0)$ changes its sign after
 crossing the critical point $s=1$. Since the product in
(\ref{2.14}) is taken over fermionic spectrum, which does not
contain the value $q=0$, this does not cause any problem there.
 However, we will see that the ambiguity in the definition of $\gamma(0)=\pm\mu$
 leads to two different representations for the correlation
 function.

 The sum over spin configurations in the right hand side of the
 expression
(\ref{2.4}) for the correlation function can also be written in
terms of pfaffians \cite{BuSha}. Corresponding matrices, however,
are not translationally invariant. This fact drastically
complicates the calculations. Nevertheless, the computation of the
correlation function can be reduced to the computation of the
determinant of a matrix
\begin{equation}
\langle\sigma(0)\sigma(\rv)\rangle=\det A^{(\mathrm
{dim})},\label{2.17}
\end{equation}
 of considerably smaller size ${\mathrm {dim}}={\mathrm
 {dim}}(\rv)$, defined by the distance between correlating spins.
 Further work is needed to transform the representation
 (\ref{2.17}) into a representation with analytic dependence
 on the distance.

Form factor representation for the correlation function of the
Ising model is the most acceptable from physical point of view.
First it was obtained in \cite{Palmev} for the infinite lattice in
the ferromagnetic region ($K>K_c$, $s>1$). Later it was extended
\cite{yamada1} to the paramagnetic case ($K<K_c$, $s<1$).
 We note that somewhat earlier a similar representation for the
 2-point Green function was deduced in \cite{Berg} via
$S$-matrix approach \cite{Zam} for a quantum field theory model
with factorized $S$-matrix ($S_2=-1$), which is usually associated
with the scaling limit of the Ising model. The discovery of the
form factor representation for the correlation function has led to
the whole trend \cite{Smirnov} in the integrable quantum field
theory.

For the finite lattice the problem seems to be more difficult, but
the result \cite{Bugrij} is, in a sense, even simpler. If
correlating spins are located along one of the lattice axes, the
matrix in the right hand side of (\ref{2.17}) has Toeplitz form.
For example, when correlating spins are located along the
horizontal axis (Fig. 1a), then
\begin{equation}\label{2.18}
\langle\sigma(\rv_1)\sigma(\rv_2)\rangle=\det A^{(|x|)},
 \qquad \rv_2-\rv_1=(x,0),
\end{equation}
and the elements of $|x|\times|x|$ matrix $ A^{(|x|)}_{k,k'}$ are
given by \cite{Bugrij}
\begin{eqnarray}&&
A^{(|x|)}_{k,k'}=\frac{1}{MN}{\sum_{\pv}}^{(f,f)}\ \frac
{\e^{ip_x(k-k')}
[2t(1+t^2)-(1-t^2)(\e^{ip_x}+t^2\e^{-ip_x})]}{(1+t^2)^2-2t(1-t^2)
(\cos p_x+\cos p_y)},\label{2.19}\\&&\ \qquad
k,k'=0,\,1,\,\ldots,\,|x|-1.\nonumber
\end{eqnarray}

 As it was shown in \cite{Bugrij} via Wiener-Hopf integral
 equations technique \cite{McCoyWu} adjusted to finite-size lattice,
 the determinant (\ref{2.18})
 can be computed analytically and for the correlation function
 one obtains
   \begin{eqnarray}\label{2.20}
\langle\sigma{(\rv_1)}\sigma{(\rv_2)}\rangle&=&(\xi\cdot\xi_T)\e^{-|x|
/\Lambda} \sum_{l=0}^{[N/2]}g_{2l}(x),\quad\hspace{1cm}{\text
{for\ \ }} \gamma(0)=\mu,\\\label{2.21}
\langle\sigma{(\rv_1)}\sigma{(\rv_2)}\rangle&=&
(\xi\cdot\xi_T)\e^{-|x|/\Lambda}
\sum_{l=0}^{[(N-1)/2]}g_{2l+1}(x),\quad{\text {for\ \ }}
\gamma(0)=-\mu,\\\label{2.22} g_n(x)&=&\frac{\e^{-n/\Lambda}}
{n!N^n}{\sum_{[q]}}^{(b)} \biggl(\prod_{i=1}^{n}
\frac{\e^{-|x|\gamma_i-\eta_i}}{\sinh \gamma_i}
\biggr)F_n^2[q],\quad g_0=1,\\\label{2.23} F_n[q]&=&
\prod_{i<j}^{n}\frac{\sin((q_i-q_j)/2)}{\sinh((\gamma_i+\gamma_j)/2)},
\quad F_1=1, \end{eqnarray} where $\gamma_i=\gamma(q_i)$,
$\eta_i=\eta(q_i)$.
   The expressions (\ref{2.20}), (\ref{2.21}) are finite sums.
   However, the upper limits of summation can be set infinite,
   since it follows from (\ref{2.23}) that the form factor
   $F_n[q]$ vanishes for $n>N$.
   Note an important detail --- the summation over the phase volume in
   (\ref{2.22}) is taken over bosonic spectrum of quasimomenta, in
   contrast with the initial fermionic spectrum, which determines the matrix
   (\ref{2.19}). The other quantities in
   (\ref{2.20})--(\ref{2.23}) are given by
\begin{eqnarray}\xi&=&|1-s^{-4}|^{1/4},\label{2.24}\\
\ln\xi_T&=&\frac{N^2}{2\pi^2}\int\limits_{0}^{\pi}\frac{dp\, dq\,
\gamma'(p)\gamma'(q)}{\sinh (N\gamma(p))\sinh
(N\gamma(q))}\ln\biggl|
\frac{\sin((p+q)/2)}{\sin((p-q)/2)}\biggr|, \label{2.25}\\
 \Lambda^{-1}&=&
\frac{1}{\pi}\int\limits_{0}^{\pi}dp\,\ln\coth({N\gamma(p)/2}),
 \label{2.26}\\
\eta(q)&=& \frac{1}{\pi}\int\limits_{0}^{\pi}\frac{dp\:(\cos p-
\e^{-\gamma(q)})}{\cosh\gamma(q)-\cos p}\ln\coth (N\gamma(p)/2).
 \label{2.27}\end{eqnarray}

 ``Cylindrical parameters'' $\xi_T$,
$\Lambda^{-1}$, $\eta(q)$  explicitly depend on the number of
sites $N$ on the base of the cylinder. Their asymptotic behaviour
for $N|\mu|\gg1$ is the following:
\begin{eqnarray}\label{2.28}
\ln\xi_T &\simeq&\frac{1}{\pi}\e^{-2N|\mu|},\\
\Lambda^{-1}&\simeq& e^{-N|\mu|}
 \sqrt{\frac{2\sinh|\mu|}{\pi N}}\label{2.29}\\
 \eta(q)&\simeq&\frac{4\e^{-N|\mu|}}{(\e^{\gamma(q)}-1)}
\sqrt{\frac{\sinh|\mu|}{2\pi N}}.\label{2.30}\end{eqnarray}
  Thus outside the critical point cylindrical parameters
 $\Lambda^{-1}$, $\ln \xi_T$ and $\eta(q)$
 exponentially decrease for large $N$ and tend to zero for infinite
 lattice.
 Finite sums (\ref{2.20}), (\ref{2.21}) transform into series,
summation over the phase volume in (\ref{2.22}) is substituted by
 integration and, as a result,  form factor representations on
the cylinder transform  into form factor representations on the
infinite lattice \cite{Palmev}, \cite{yamada1}. For any finite $N$
both expansions --- over even $n$ (\ref{2.20}) and over odd $n$
(\ref{2.21}) --- are valid in both ferromagnetic $(s>1)$ and
paramagnetic $(s<1)$ regions. However, for $N\rightarrow \infty$,
the first series is well-defined and converges in the
ferromagnetic region and the second one does so in the
paramagnetic region.

 Recall that we started from the determinant (\ref{2.18})
 of a $|x|\times|x|$ matrix. The number of terms in its formal
 definition rapidly increases when $x$ grows.
 However, the form factor representations (\ref{2.20})--(\ref{2.23})
 are  finite sums for any fixed $N$, and the number of terms in
 these sums
 does not depend on $|x|$. This gives a unique opportunity to
 verify (\ref{2.20})--(\ref{2.23}) by  comparing these
 representations with
 the results of transfer matrix calculations for $N$-row Ising chains.
 For fixed $N$ the dimension of the corresponding transfer matrix
 is equal to $2^N\times2^N$.
 One can find analytically all eigenvectors and eigenvalues
 if $N$ is not too large. We have successfully performed such check analytically for
 $N=2,3,4$ and numerically --- for
$N=5,6$.

\section{Correlation function $\langle\sigma(0,0)\sigma(x,y)\rangle$}
 The rigorous derivation of the form factor
 representation on the cylinder was performed
 in \cite{Bugrij} only for the spins located along the cylinder
 axis.
 We have not yet succeeded in generalization of the method for
 arbitrary disposition of correlating spins (Fig. 1b). Meanwhile,
 the calculation of the momentum representation of correlation
 function
\begin{equation}
\label{3.1}
\widetilde{G}(\pv)=\sum_{\rv}\e^{i\pv\rv}\langle\sigma(0)\sigma(\rv)
\rangle,
\end{equation}
 or the susceptibility (which is related to
$\widetilde{G}(\pv=0)$) requires an explicit dependence on both
components of the vector $\rv$. Form factor representations
~(\ref{2.20})--(\ref{2.23}) have a transparent physical content.
 This allows to make reasonable assumptions for corresponding
 generalizations. The above mentioned possibility of independent check
  allows to eliminate wrong hypotheses and to make correct choice.
  In principle, when $y$-component of the vector $\rv$ is not
  equal to
  zero, all quantities in (\ref{2.20})--(\ref{2.23}) could change their
  form. However, corresponding expressions for free bosons and fermions on the
  lattice prompt one of the simplest generalizations --- just the substitution
   $$ \e^{-|x|\gamma(q)}\to\e^{-|x|\gamma(q)-iyq}. $$
 Suprisingly enough, this turns out to be sufficient. If instead of
$g_n(x)$~(\ref{2.22}) one uses the expression
\begin{equation}
g_n(\rv)=\frac{\e^{-n/\Lambda}}{n!N^n}{\sum_{[q]}}^{(b)}\prod_{j=1}^n
\left(\frac{\e^{-|x|\gamma_j-iyq_j-\eta_j}}{\sinh\gamma_j}\right)F_n^2[q],
\quad g_0=1, \label{3.3}
\end{equation} then the correlation functions~(\ref{2.20}) and (\ref{2.21})
exactly coincide with the transfer matrix results for $N=2,3,4$ in
the whole range of the variables $x$, $y$, $K$. Numerical
calculations confirm this for $N=5,6$ as well. The validity of
~(\ref{3.3}) is out of doubts and we hope that the known answer
will simplify the problem of its rigorous derivation.

As an illustration, consider the example of $N=3$. The expansions
(\ref{2.20})--(\ref{2.21}) are very similar to the representation
of the correlation function in terms of  the transfer matrix
eigenvalues
\begin{equation}
\langle\sigma(0)\sigma(\rv)\rangle=a_1(y)(\lambda_1/\lambda_0)^{|x|}+
a_2(y)(\lambda_2/\lambda_0)^{|x|}+\cdots,\label{3.4}
\end{equation}
 where $\lambda_0$ is the largest eigenvalue and the coefficients
$a_j(y)$ are given by some bilinear combinations of the components
of eigenvectors. To reduce, for example, (\ref{2.21}) to
(\ref{3.4}), we use the following expressions for the cylindrical
parameters $\xi_T$, $\Lambda^{-1}$, $\eta(q)$:
 \begin{eqnarray}\label{3.5}\Lambda^{-1}&=&\frac12
 \biggl({\sum_q}^{(f)}\gamma(q)
 -{\sum_q}^{(b)}\gamma(q)\biggr),\\
 \label{3.6}
 \e^{-\Lambda^{-1}-\eta(q_i)}&=&\frac{{\prod\limits_q}^{(b)}
\sinh\left( \frac{\gamma(q)+\gamma(q_i)}{2}\right)}
{{\prod\limits_q}^{(f)}\sinh\left(
\frac{\gamma(q)+\gamma(q_i)}{2}\right)},\\
\label{3.7}\xi_T^4&=&\frac{{\prod\limits_q}^{(b)}{\prod\limits_p}^{(f)}\sinh^2\left(
\frac{\gamma(q)+\gamma(p)}{2}\right)}
{{\prod\limits_q}^{(b)}{\prod\limits_p}^{(b)}\sinh\left(
\frac{\gamma(q)+\gamma(p)}{2}\right){\prod\limits_q}^{(f)}{\prod\limits_p}^{(f)}
\sinh\left( \frac{\gamma(q)+\gamma(p)}{2}\right)}. \end{eqnarray}
 One can derive these expressions from (\ref{2.25})--(\ref{2.27})
 by passing to contour integrals in the variable $z=\e^{iq}$ and
 computing the residues.

For $N=3$ we have from (\ref{3.5})--(\ref{3.7}) and (\ref{2.24})
\begin{eqnarray}\label{3.8}
\Lambda^{-1}&=&\frac{1}{2}\bigl[
 \gamma(\pi)+2\gamma({\pi}/{3})-\gamma(0)-2\gamma(
 {2\pi}/{3})
 \bigr],\\
 \xi\xi_{T}&=&\frac
 {\sinh\frac{\gamma(0)+\gamma(\pi/3)}{2}
  \sinh\frac{\gamma(\pi)+\gamma(2\pi/3)}{2}
  \sinh^{2}\frac{\gamma(2\pi/3)+\gamma(\pi/3)}{2}}
 {\sinh\frac{\gamma(0)+\gamma(2\pi/3)}{2}
  \sinh\frac{\gamma(\pi)+\gamma(\pi/3)}{2}
  \sinh\gamma(\pi/3)\sinh\gamma(2\pi/3)},\\
e^{-\Lambda^{-1}-\eta(q)}&=&
 \frac
 {\sinh\frac{\gamma(0)+\gamma(q)}{2}
  \sinh^{2}\frac{\gamma(2\pi/3)+\gamma(q)}{2}}
 {\sinh\frac{\gamma(\pi)+\gamma(q)}{2}
  \sinh^{2}\frac{\gamma(\pi/3)+\gamma(q)}{2}}.\label{3.10}
  \end{eqnarray}
 Then one finds
\begin{eqnarray}\label{3.11}
\ln(\lambda_0/\lambda_1)&=&\Lambda^{-1}+\gamma(0),\\
\ln(\lambda_0/\lambda_2)&=&\Lambda^{-1}+\gamma(2\pi/3),\\
\ln(\lambda_0/\lambda_3)&=&\Lambda^{-1}+\gamma(0)+2\gamma(2\pi/3),
\end{eqnarray}
\begin{eqnarray}
\label{3.14}
 a_1(y)&=&\frac{1}{3} \frac
 {\sinh\frac{\gamma(0)+\gamma(2\pi/3)}{2}
  \sinh\frac{\gamma(\pi)+\gamma(2\pi/3)}{2}
  \sinh^{2}\frac{\gamma(2\pi/3)+\gamma(\pi/3)}{2}}
 {\sinh\frac{\gamma(0)+\gamma(\pi/3)}{2}
  \sinh\frac{\gamma(\pi)+\gamma(\pi/3)}{2}
  \sinh\gamma(\pi/3)\sinh\gamma(2\pi/3)},\\ a_2(y)&=&\frac{2}{3}\frac
 {\sinh\frac{\gamma(0)+\gamma(\pi/3)}{2}
  \sinh\frac{\gamma(0)+\gamma(\pi)}{2}}
 {\sinh\gamma(\pi/3)
  \sinh\frac{\gamma(\pi/3)+\gamma(\pi)}{2}} \cos
  (2\pi y/3),
\\ \label{3.16} a_3(y)&=&\frac{1}{64}\frac
  {1}
  {\sinh\frac{\gamma(0)+\gamma(\pi/3)}{2}
  \sinh\frac{\gamma(\pi)+\gamma(\pi/3)}{2}
  \sinh\frac{\gamma(0)+\gamma(2\pi/3)}{2}
  \sinh\frac{\gamma(\pi)+\gamma(2\pi/3)}{2}
    } \times\\
    \nonumber&& \times\frac{1}{\sinh\gamma(\pi/3)\sinh\gamma(2\pi/3)
  \sinh^{2}\frac{\gamma(\pi/3)+\gamma(2\pi/3)}{2}}.\end{eqnarray}

 Our $2^3\times 2^3$ transfer matrix  has 8 eigenvalues, and some of them are
 equal. Besides that, some eigenvectors have zero components.
As a result, the expression for the correlation function
(\ref{3.4}) contains
 only three (not seven) independent terms.
 If we take into account the definition (\ref{2.15}), (\ref{2.16})
 of the function $\gamma(q)$ for particular values of quasimomentum
 $q=0,\ \pi/3,\ 2\pi/3,\
\pi$, we get an exact correspondence between these three terms and
(\ref{3.11})--(\ref{3.16}).

\section{Momentum representation of the correlation\\ function}

Since we have the expression (\ref{3.3}) for $g_n(\rv)$, which
explicitly depends on both components of $\rv$, we can make the
Fourier transform. Let us write the momentum representation of
(\ref{3.1}) in a form similar to (\ref{2.20})--(\ref{2.21}):
\begin{equation}\label{4.1}
\widetilde{G}(\pv)=\xi\xi_T\sum_n\widetilde{g}_n(\pv),\end{equation}
\begin{equation}\label{4.2}
\widetilde{g}_n(\pv)=\sum_{\rv}\e^{-|x|/\Lambda}g_n(\rv)\e^{i\pv\rv},
\end{equation}
where
\begin{equation}
\sum_{\rv}=\sum_{x=-\infty}^\infty\sum_{y=1}^N.
\end{equation}
Performing the summation in (\ref{4.2}), we find
\begin{equation}\label{4.4}
\widetilde{g}_n(\pv)=\frac{\e^{n/\Lambda}}{n!N^{n-1}}
{\sum_{[q]}}^{(b)}\biggl(\prod_{j=1}^n\frac{\e^{-\eta_j}}{\sinh\gamma_j}
\biggr)\frac{\sinh\biggl(\Lambda^{-1}+\sum\limits_{j=1}^n\gamma_j
\biggr)F_n^2[q]}
{\cosh\biggl(\Lambda^{-1}+\sum\limits_{j=1}^n\gamma_j\biggr)-\cos
p_x}\delta\biggl(p_y-\sum_{j=1}^nq_j\biggr).
\end{equation}
The $x$-component of the quasimomentum has a continuous spectrum,
$p_x\in[-\pi,\pi]$, but  $p_y$ is discrete:
 $$ p_y=\frac{2\pi
l}{N},\qquad l=1,\,2\,\,\ldots\,N. $$
 Corresponding
$\delta$-function in the right hand side of (\ref{4.4}) has the
meaning of the Kronecker symbol
 $$
\delta\biggl(p_y-\sum_{j=1}^nq_j\biggr)=\delta\biggl(l-\sum_{j=1}^nl_j
\biggr)\biggm|_{\displaystyle{{\rm{mod}}\,N}}\,. $$ The function
$\widetilde{g}_n(\pv)$ is periodic in $p_x$, $p_y$ with the period
$2\pi$. Inserting the ``unity''  $$
1=\int\limits^{\Lambda^{-1}+n\gamma(\pi)}_{\Lambda^{-1}+n\gamma(0)}d\omega
\,\delta\biggl(\Lambda^{-1}+\sum^n_{j=1}\gamma_j-\omega\biggr), $$
into the sum (\ref{4.4}) (here $\delta$ denotes Dirac
$\delta$-function) and interchanging the order of summation and
integration, we obtain
\begin{equation}\label{4.8}
\widetilde{g}_n(\pv)=\int\limits^{\Lambda^{-1}+n\gamma(\pi)}
_{\Lambda^{-1}+n\gamma(0)}d\omega\frac{\sinh\omega}{\cosh\omega-\cos
p_x}\rho_n(\omega, p_y),
\end{equation}
\begin{equation}\label{4.9}
{\rho}_n(\omega,p_y)=\frac{\e^{-n/\Lambda}}{n!N^{n-1}}
{\sum_{[q]}}^{(b)}\biggl(\prod_{j=1}^n\frac{\e^{-\eta_j}}{\sinh\gamma_j}
\biggr)F^2_n[q]\delta\biggl(\Lambda^{-1}+\sum_{j=1}^n\gamma_j-
\omega\biggr)\delta\biggl(p_y-\sum_{j=1}^nq_j\biggr).\end{equation}
On the infinite lattice in the scaling limit the rotational
symmetry is restored and (\ref{4.8}), (\ref{4.9}) transform into
the classical Lehmann representation in the quantum field theory.

\section{Magnetic susceptibility}
 On $M\times N$ square lattice with equal horizontal and vertical
 coupling parameters the partition function $Z$ depends on four
 variables
\begin{equation}
Z=Z(K,h,N,M)=\sum_{[\sigma]}\e^{-\beta
H[\sigma]+h\sum\limits_{\rv}\sigma(\rv)},
\end{equation}
where dimensionless parameter $h=\beta \HC$,
 $\HC$ --- magnetic field.
The magnetization $\M$ and magnetic susceptibility $\chi$ can be
expressed in terms of the derivatives of the partition function
with respect to $h$:
\begin{equation}
\M(K,h,N,M)=\frac{1}{MN}\frac{\partial\ln Z}{\partial
h}=\langle\sigma\rangle,
\end{equation}
\begin{equation}\label{5.3}
\beta^{-1}\chi(K,h,N,M)=\frac{\partial\M}{\partial
h}=\sum_{\rv}\biggl(\langle\sigma(0)\sigma(\rv)\rangle-\langle
\sigma\rangle^2\biggr).
\end{equation}
The magnetization for $h=0$ and finite $M$, $N$ is equal to zero
due to $\mathbb{Z}_2$-symmetry of the Ising model. This holds even
when one of the dimensions is set infinite. In the last case,
when, for example, $M\to\infty$, $N=\const$, 2D Ising model
transforms into a 1D chain with $N$ rows, for which the
spontaneous symmetry breaking is impossible. The susceptibility
can be easily computed from (\ref{4.1})--(\ref{4.4})
   \begin{eqnarray}\label{5.4}
\chi&=&\chi_0+ \sum_{l=1}^{[N/2]}\chi_{2l}\ {\text{\hspace{2.3cm}
 for\ \ }} \gamma(0)=\mu,\\\label{5.5}
\beta^{-1}\chi_0&=&\xi\xi_TN\coth(1/2\Lambda),\\\label{5.6}
\chi&=&\sum_{l=0}^{[(N-1)/2]}\chi_{2l+1}\ {\text{\hspace{2.3cm}
for\ \ }}\gamma(0)=-\mu,\\\label{5.7}
\beta^{-1}\chi_n&=&\frac{\e^{-n/\Lambda}}{
n!N^{n-1}}{\sum_{[q]}}^{(b)} \biggl(\prod_{i=1}^{n}
\frac{\e^{-\eta_i}}{\sinh \gamma_i}
\biggr)F_n^2[q]\coth\biggl[\frac12\biggl(\Lambda^{-1}+\sum_{i=1}^n
\gamma_i\biggr)\biggr]\delta\biggl(\sum_{i=1}^n q_i \biggr).
\end{eqnarray}
In the paramagnetic region $(s<1)$ the expression (\ref{5.6})
admits the limit $N\to\infty$ and tends to the susceptibility on
the infinite lattice. However, in the ferromagnetic region $(s>1)$
one can consider the limit $N\to\infty$ only for the quantity
\begin{equation}\label{5.8}
\chi_F=\chi-\chi_0=\sum_{l=1}^{\infty}\chi_{2l},\end{equation}
which reproduces well-known zero-field ferromagnetic
susceptibility of the Ising model in thermodynamic limit. For
large but finite $N$ the main contribution to the susceptibility
is given by the term $\chi_0$
\begin{equation}
\beta^{-1}\chi_0\simeq2\xi N\Lambda\simeq\frac{\sqrt{\pi}\xi
N^{3/2}}{\sqrt{\sinh|\mu|}}\e^{N|\mu|},\label{5.9}
\end{equation}
which exponentially increases with the growth of the size of the
cylinder base. It follows from (\ref{5.9}) that the larger $N$ ---
the smaller field $ \delta h\sim\e^{-N|\mu|}$ is needed to order
all spins on the lattice.

 Unfortunately, the exact solution for the partition function of
 the Ising model in external field is not known. However, the appearance
 of spontaneous magnetization can be deduced
 from the analysis of high- and low-temperature expansions.
 The rigorous definition of spontaneous magnetization is given by
 the following order of limits according to the Bogolyubov concept
 of quasiaverages:
\begin{equation}
\M_0(K)=\lim_{h\to0}\bigl[\lim_{M,N\to\infty}\M(K,h,N,M)\bigr].
\end{equation}
 However, if we conjecture the decreasing of correlations at large
 distances and the possibility of interchanging of the corresponding
 limits, we can find exact solution for the squared
 spontaneous magnetization. It is equal to spin-spin correlation
 function (\ref{2.24}) with infinite distance between correlating spins
\begin{equation}\label{5.12}
\M^2_0(K)=\lim_{|\rv|\to\infty}\langle\sigma(0)\sigma(\rv)\rangle=
\langle\sigma(0)\rangle\langle\sigma(\infty)\rangle=
\langle\sigma\rangle^2=\xi.
\end{equation}
Meanwhile, the sums over lattice of each summand in the right hand
side of (\ref{5.3}) do not converge in the thermodynamic limit.
Therefore, the substitution of $\M^2(K,0,\infty,\infty)$ by the
limiting value of the correlation function (which equals $\xi$)
under the (infinite) sum in the last step of the limits $h\to0$,
$M,N\to\infty$ requires not only (\ref{5.12}), but also the
existence of the limit
\begin{equation}\label{5.13}
\lim_{h\to0}\bigl\{\lim_{M,N\to\infty}MN\bigl[\M^2(K,h,M,N)
-\xi\bigr]\bigr\}=f(K),
\end{equation}
and, moreover,
\begin{equation}\label{5.14}
 f(K)=0.
 \end{equation}
The explicit dependence of the correlation function on the size
$N$, namely, the exponential tending of cylindrical parameters to
their limiting values (\ref{2.28})--(\ref{2.30}), can be viewed as
an argument in favor of the equalities (\ref{5.13}), (\ref{5.14}).

The behaviour of the correlation function at large distances in
the ferromagnetic region is mainly determined by the first term in
the expansion (\ref{2.20}). Note that it does not depend on
$y$-projection of $\rv$
\begin{equation}
G_0(|\rv|)=\xi\xi_T\e^{-|x|/\Lambda}.
\end{equation}
 Therefore, the distance $\sim\Lambda$, for which spins are strongly
 correlated, rapidly increases (cf. with (\ref{2.29})) with the growth of $N$.
 Physically it means that for ``ferromagnetic'' temperatures the cylinder
 is divided into ``domains'' of size $\sim\Lambda$ with nonzero
 magnetization, the magnetization of the whole infinite cylinder
 being equal to zero. It is clear that the squared spontaneous
 magnetization would be more naturally defined by the value of the
 correlation function at large distances $|\rv|=R(N)$,
 which do not exceed the size of the domain
 $$ N\ll R(N)\ll\Lambda.$$
 It follows from (\ref{2.29}) that for sufficiently large $N$
these inequalities can be satisfied. In accordance with this, the
sum over $x$ with infinite limits in the definition of the
thermodynamic limit of the susceptibility (\ref{5.3}) should be
substituted by a sum with the limits that do not exceed the size
of the domain. In this case the condition
 $$
\sum_{x=-R}^R\sum_{y=1}^N[G_0(|\rv|)-G_0(R)]\simeq\xi
NR^2/\Lambda\mathop{\to}_{N\to\infty}0,$$ can be treated as a
formal substantiation of the definition (\ref{5.8}) of the
susceptibility in the ferromagnetic phase. We can now estimate the
``thermodynamic cutoff parameter'' $R(N)$: $$
R(N)\ll\sqrt{\Lambda/N\xi}\simeq\e^{N|\mu|/2}[\pi/(2N\sinh|\mu|)]^{1/4}.
$$ We believe that these estimates slightly clarify the physical
content of the formal thermodynamic limit procedure.

\section{Singularity structure}
 The initial expression (\ref{2.3}) for the partition function of
 the Ising model is a polynomial in
 $s$, and the solution (\ref{2.12}) is the factorized form of this
 polynomial.
 It provides an example of the mechanism of Lee-Yang ``zeros''
 \cite{Lee}, which stipulates the appearance of critical singularities
 in the thermodynamic limit. The roots of the polynomial (\ref{2.12})
are located on the unit circle $|s|=1$ in the complex $s$ plane.
For any finite $M$ and $N$ the zero $s=1$ on the real axis  does
not appear, since the fermionic spectrum does not contain the
value of quasimomentum $q_x=q_y=0$. When one of the dimensions
increases then zeros are concentrated on the circle $|s|=1$,
forming a dense set. In the limit $M\to\infty$, $N=\const$ they
are transformed into a finite number (equal to $N$) of the root
type branchpoints, located on the circle $|s|=1$. To see this, one
has to use the representation (\ref{2.14}) and the definition
(\ref{2.15}), (\ref{2.16}) of the function $\gamma(q)$. These
branchpoints, in turn, form a dense set with the growth of $N$,
but in the limit $N\to\infty$ they are transformed into four
isolated logarithmic branchpoints $s=\pm1,\,\pm i$. As a result,
the specific heat in the thermodynamic limit acquires a
logarithmic divergence $\sim\ln|1-s|$. It is worth noticing that
the specific heat is expressed through the same function in both
ferromagnetic and paramagnetic regions of $s$.

One could think that a similar picture holds for susceptibility.
Indeed, the initial expression (\ref{2.4}) for the correlation
function for finite $M$ and $N$ is a ratio of polynomials in $s$.
The formation of the singularities of the partition function,
which stands in the denominator, we have just briefly described.
Unfortunately, the polynomial in the numerator cannot be written
in such simple factorized form. Nevertheless, our form factor
representation for $M\to\infty$ and finite $N$ shows that the
correlation function has a finite number of root branchpoints on
the circle $|s|=1$. Their number is doubled in comparison with the
case of partition function, since the expressions
(\ref{2.20})--(\ref{2.23}), (\ref{3.3}) contain the functions
$\gamma(q)$ (\ref{2.15}), corresponding to both bosonic and
fermionic values of quasimomentum. The susceptibility on the
cylinder is given by the infinite sum of correlation functions and
this can lead to the appearance of additional singularities. One
can show, however, that these singularities do not appear on the
first sheet of the appropriate Riemann surface.

 As an example, let us write down the susceptibility
$\chi$ (\ref{5.4}) for $N=3$, using the expressions
(\ref{3.14})--(\ref{3.16}) and representations
(\ref{3.8})--(\ref{3.10}) for cylindrical parameters
 \begin{eqnarray}\nonumber\beta^{-1}\chi&=&\frac
 {\sinh\frac{\gamma(0)+\gamma(2\pi/3)}{2}
  \sinh\frac{\gamma(\pi)+\gamma(2\pi/3)}{2}
  \sinh^{2}\frac{\gamma(2\pi/3)+\gamma(\pi/3)}{2}}
 {\sinh\frac{\gamma(0)+\gamma(\pi/3)}{2}
  \sinh\frac{\gamma(\pi)+\gamma(\pi/3)}{2}
  \sinh\gamma(\pi/3)\sinh\gamma(2\pi/3)}\:
  \coth\left(\tfrac{\Lambda^{-1}+\gamma(0)}{2}\right)+ \\
&& +\frac{1}{64}\frac{1}
  {\sinh\frac{\gamma(0)+\gamma(\pi/3)}{2}
  \sinh\frac{\gamma(\pi)+\gamma(\pi/3)}{2}
  \sinh\frac{\gamma(0)+\gamma(2\pi/3)}{2}
  \sinh\frac{\gamma(\pi)+\gamma(2\pi/3)}{2}
    } \times\label{6.1}\\
    \nonumber&& \times\frac{1}
     {\sinh\gamma(\pi/3)\sinh\gamma(2\pi/3)
  \sinh^{2}\frac{\gamma(\pi/3)+\gamma(2\pi/3)}{2}}
  \,\coth\left(\tfrac{\Lambda^{-1}+\gamma(0)+2\gamma(2\pi/3)}{2}\right).
 \end{eqnarray}
 The singularities in $s$ could appear due to zero denominator in (\ref{6.1}).
 It is easily seen, however, that the corresponding factors
 $$\sinh\frac{\gamma(q)+\gamma(q')}{2}=(\cos q'-\cos q)/\sinh\frac
 {\gamma(q)-\gamma(q')}{2}$$
 for $q\neq q'$ are not equal to zero. It can also be shown that on the first
 sheet of  Riemann surface (which is determined by the condition of
 positivity of $\gamma(q)$, treated as functions of $s$, for real $s>0$)
 the arguments of cotangents in
 (\ref{6.1}) also have non-zero values: these factors appear as the result
 of summation over coordinate $x$. Therefore, the complete set of
  singularities of the susceptibility is exhausted by the
 branchpoints contained in the functions
 \begin{equation}
 \e^{\gamma(q)}=\biggl[\sqrt{\frac12(s+s^{-1})+\sin^2\frac q2}+
\sqrt{\frac12(s+s^{-1})-\cos^2\frac q2}\biggr]^2\label{6.2}
 \end{equation}
For each value of quasimomentum $q\neq0, \pi$ the function
(\ref{6.2}) has four branchpoints. If we denote them by
$s_c=|s_c|\e^{\pm i\varphi_c}$, then
\begin{equation}
|s_c|=1, \quad
\cos\varphi_c=\left\{\begin{array}{r}\cos^2q/2\\-\sin^2q/2\end{array}
\right.\,.\label{6.3}
\end{equation}
It is seen from (\ref{6.2}), that for $q=0, \pi$ there exist only
two branchpoints $s_c=\pm i$. One can now show that for any fixed
$N$ the total number of singularities is equal to $4N-2$, and all
singularities are located on the unit circle $|s|=1$. We represent
the corresponding picture for $N=3$ in the Fig. 2.
\begin{figure}[h] \begin{center}
 \includegraphics[height=50mm,keepaspectratio=true]
 {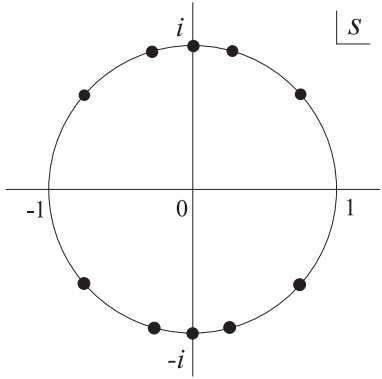}
 \caption{The location of the singularities of susceptibility
 $\chi$
 in the complex plane $s=\sinh 2\beta J$ for $N=3$.}
 \end{center} \end{figure}
  We do not discuss the limit $N\to\infty$, when the singularities on
  the circle $|s|=1$ form a dense set.
  This problem was seriously analyzed in
\cite{Nickel1}--\cite{Nickel2}, where the authors conjecture that
the singularities form a natural boundary $|s|=1$ for the
susceptibility.

 \vspace{1cm}
 We thank V. N. Shadura for his assistance in our work and helpful
 discussions;
 we are also grateful to Professor B. M. McCoy for useful
 comments on the form factor representation of the correlation function
 and for bringing the problem of singularities
 of the susceptibility to our attention.

 This work was supported
 by the INTAS program under grant INTAS-00-00055.

  \end{document}